Emotion Recognition of the Singing Voice:

Toward a Real-Time Analysis Tool for Singers

Daniel Szelogowski

# Abstract

Current computational-emotion research has focused on applying acoustic properties to analyze how emotions are perceived mathematically or used in natural language processing machine learning models. While recent interest has focused on analyzing emotions from the spoken voice, little experimentation has been performed to discover how emotions are recognized in the singing voice – both in noiseless and noisy data (i.e., data that is either inaccurate, difficult to interpret, has corrupted/distorted/nonsense information like actual noise sounds in this case, or has a low ratio of usable/unusable information). Not only does this ignore the challenges of training machine learning models on more subjective data and testing them with much noisier data, but there is also a clear disconnect in progress between advancing the development of convolutional neural networks and the goal of emotionally cognizant artificial intelligence. By training a new model to include this type of information with a rich comprehension of psycho-acoustic properties, not only can models be trained to recognize information within extremely noisy data, but advancement can be made toward more complex biofeedback applications – including creating a model which could recognize emotions given any human information (language, breath, voice, body, posture) and be used in any performance medium (music, speech, acting) or psychological assistance for patients with disorders such as BPD, alexithymia, autism, among others. This paper seeks to reflect and expand upon the findings of related research and present a stepping-stone toward this end goal.

## 1. Introduction

While speech-emotion analysis research has become more prominent in recent years, there has been an overwhelming lack of study on emotional analysis of singing voices. Most current research is aimed toward classifying an audio clip of one emotion and has been primarily focused on pre-recorded audio clips of spoken (and sung) voices, none providing real-time feedback and audio-sectional descriptors. Recent focus has been tested on pure audio clips, none attempting to analyze noisy (especially accompanied voice) data. Ideally, this research will serve as the backend architecture for a real-time analysis tool or mobile app for singers and teachers to gauge the emotional valence and activation of various points within a piece of music, either through pre-recorded audio or live feedback through the utilization of machine learning. Visual feedback, or **biofeedback**, has been scientifically proven to strengthen "mind-to-motor" coordination and help to influence production toward improvement[1] – through the implementation of a visual analysis tool for recognizing sung emotions in real-time, such a tool would be extremely useful for musicians in training and professionals alike, and — eventually — for full ensembles.

Emotional expression is a challenge well-known to actors and musicians alike, especially for singers of any sort. While the portrayal of emotions in music is highly subjective, musicians must have a thorough interpretation in mind before performing for an audience if they seek to truly exhibit the characterization of the music; given that musical harmony alone can portray an expression, this generally coincides with the setting of the text of the piece. As such, the voice alone exhibits many features which express various emotions: the breath and its fullness or depth, the volume, pressure, and stability of the voice, among others. While reading these expressions is instinctual for most people, replicating the same ability in artificial intelligence is challenging.

## 1.1 Machine Learning and Training

The most likely candidate for building an emotion-detecting system in the field of machine learning is the **neural network (NN)** – a form of artificial intelligence that seeks to replicate biological learning in the brain through the development of neurons and synaptic connections. One specialized form of NN, known as a **Convolutional Neural Network (CNN)**, is a type of deep-learning neural network primarily applied to analyzing visual imagery for image/video recognition and classification, and natural language processing.[2] This type of artificial intelligence is inspired by biological processes, wherein the connectivity pattern between neurons resembles the organization of the animal visual cortex.[3] To train these types of networks, a large, consistent dataset is especially necessary for creating a reliable and accurate prediction and classification system. One such dataset is the **Ryerson Audio-Visual Database of Emotional Speech and Song (RAVDESS)**: a database of 24 professional actors (12 male and female, each) containing audio and video recordings of 8 different emotions (neutral, calm, happy, sad, angry, fearful, disgust, surprise) demonstrated both spoken and sung, primarily used to study the relationship between face and voice expressions.[4]

## 2. Related Work

While the topic of sung emotion recognition has not been discussed frequently in recent research, a great amount of experimentation and analysis has been performed on speech emotion recognition – not only from a machine learning perspective but also purely scientific through acoustic and psychological studies with **valence** (intrinsic attractiveness/aversiveness) and **activation** (arousal, stimulation response) analysis. **Spectral analysis** – a type of visual-based audio analysis used in measuring the distribution of acoustic energy across frequencies[5] – has become increasingly useful in the fields of both singing and speech pathology alike – it provides a means of either analyzing energy distribution in speech/sung sound (through the **Fast Fourier Transform, FFT**), including voice harmonics, or estimating the vocal tract filter that shaped the sound (**Linear Predictive Coding, LPC**).[6] As such, spectrographic tools may be utilized as a means of biofeedback to aid in the teaching of voice-related studies, especially for visual learners (see Figure 1).[7] By applying the inverse FFT to an audio signal, however, we are able to return the **cepstrum** of the spectral data, or "spectrum of a spectrum" – a means of analyzing sound and vibration of a signal – allowing us to measure how the power of the signal is distributed over frequency and its density through the **power cepstrum**.[8] More precisely, the **Mel-frequency cepstrum (MFC)** can be used to represent this data on a short-term basis[9] from the Mel scale, a scale relating a sound's perceived frequency to its actual frequency, which can be used to scale a sound close to how it would be perceived by the human ear.[10]

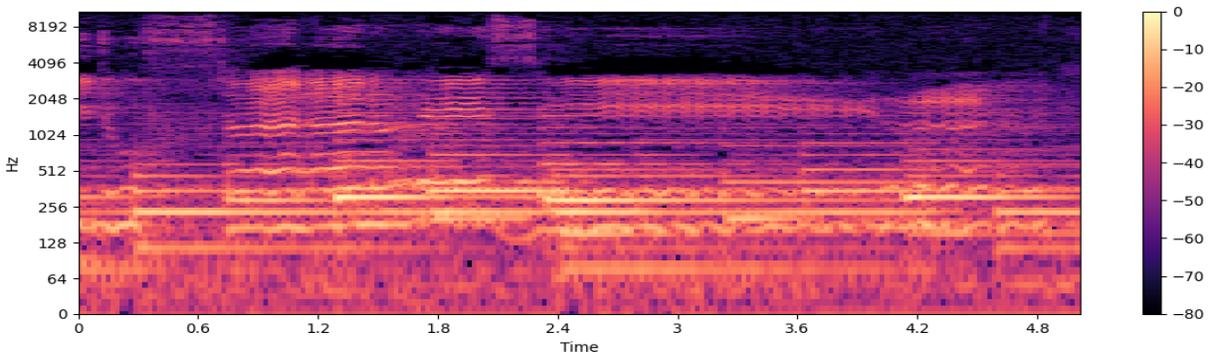

**Figure 1:** Spectral analysis of a baritone voice singing "Allerseelen" by Richard Strauss with piano accompaniment across five seconds (see *Appendix A, Section 1*). The "hot" waves on the graph, seen especially between 128-1024 Hz, represent the vocal frequencies in the recording.

## 2.1 Speech Emotion Recognition and Mel-Frequency Cepstral Coefficients

Recognizing emotions in normal speech is a process carried out by the amygdala, a region of the brain within the medial temporal lobe involved in emotional processes as part of the limbic system: the body's own neural network that handles various aspects of memory and emotion.[11] The amygdala responds to two properties of emotions: valence, measured as positive (attractive) or negative (aversive), and intensity, measured from low to high.[12] Thus, a CNN is the most effective machine learning model to represent the amygdala given its classification ability. To recreate the analytical process of the amygdala within the CNN, one way to measure the emotional properties of sound is through the coefficients of the MFC measurements of an audio clip, known as **Mel-frequency cepstral coefficients (MFCCs)** which also assists in better representing compressed audio on the Mel scale.[13]

### 2.1.1   Emotional Expression in Performance

For singers, performing expressively through the voice is a challenging form of acting – especially for languages foreign to the vocalist; 'confidence' alone is insufficient.[14] Miller discusses:

> Suppose [that] the teacher is dealing with a singer who has a good voice and considerable technical facility, but [finds] difficulty in achieving communication out of fear of "looking silly" in public. […] It sometimes occurs that the outgoing tendencies a person displays in everyday social encounters are not necessarily transferable when that person turns to the field of performance. On the other hand, the reserved personality may more readily find a communicative channel through the musical setting of text and the drama than does the demonstrative person. But most often, special attention must be taken to awaken the taciturn person to his or her potential for improving communication skills in performance.[15]

Musical characterization, which often includes some physical movement, creates a sort of secondary reality where the actor is freed from behavioral responsibility and thus embodies the character of the text.[16] In a cycle of repertoire, such as an individual recital or ensemble, however, the character often changes drastically either within more complex songs or between pieces – this challenge alone is the goal of this research. As such, singers may utilize various techniques to express the desired character – which the music often assists with: stronger emotions such as anger are often expressed through an increase in sound volume and faster, often agitated rhythms – thus, a 'stronger' or 'angrier' sound – while weaker emotions are most often seen concurrent to more somber, quiet, and slower music.[17] Other factors may also be taken into account, such as the depth of the breath, the width of the singer's vibrato, the range of musical pitches (i.e., higher or lower), among vocal timbre or 'color' and other vocal effects or tones (whispered, 'hooty', harsh, brassy, etc.).[18]

## 2.2 Vocal Isolation

One aspect that may potentially affect the outcome of the CNN is the noise of the data it trains and/or tests on. In the case of this research, the training data is pure audio containing raw vocals with no noise[19] – this will help to keep the information received from the MFCCs of the audio clean and reliable; however, this is an unrealistic expectation for a biofeedback app and is difficult to achieve from a field scenario. Vocal isolation continues to be an ongoing battle of its own, especially for noisy data such as an accompanied voice (even with complex machine learning algorithms),[20] although simple solutions using FFT and **soft masking** (separating the desired information from noisy data using a threshold mask) appear viable for the time being, especially given that FFT replicates how the ear transforms audio information into sound.[21]

## 2.3 Sung Emotion Color and Recognition

Recognizing emotional expression and coloring has also proven to be equally, if not more challenging than vocal isolation. Recent attempts have included analyzing the acoustic features of a vocal signal, including **low-level descriptors** (**LLDs**, data features closely related to the origin signal/sound), **delta (difference) coefficients**, and aggregation of **long-term average spectrum-based features** (or **LTAS**, features that provide means to view the average power cepstrum over time),[22] rather than training a NN model – though this research proved an increased correlation between MFCCs and emotion recognition. Further acoustic research showed a strong correlation in emotional expression with additional acoustic parameters including proportion energy below 500Hz and 1000Hz, alpha ratio (sound absorption), spectral flatness, Hammarberg index (quantification of difference in volume between high and low speech frequencies), maximum flow declination rate (or MFDR, regarding vocal intensity in the glottis) especially, among others.[23] Lastly, another recent approach was performed by the measurement of Signal-to-Noise Ratio (SNR) levels and their correlation to emotional valence, comparing mathematically predicted emotion values to actual perceived emotions in humans.[24]

## 3. Approach

The backend system presented in this research will consist of three primary components: the CNN model, a vocal isolator, and a **WAV-file** audio recorder/splitter. The CNN will be trained using the RAVDESS dataset (see *Section 1.1*) which contains 1,012 sung text examples across the various actors, expressing one of the following emotions per clip: neutral, calm, happy, angry, sad, or fear. The WAV-file system will simply record the user's microphone as a WAV file and feature a system to divide an audio file into user-defined seconds-long segment files for evaluation purposes as well as in testing real-time biofeedback (such as recording the user's voice and providing new feedback every N-seconds), both as live and historical data. Lastly, the vocal isolator will use the spectral data obtained from an audio file to separate the voice from any accompaniment or background noise through FFT and **Wiener filtering** (see *Section 2.2*).[25] Accuracy is an issue regardless of training due to the lack of datasets intentionally designed for such testing, but approximately 75% accuracy is a reasonable goal in this instance.

## 4. Implementation

The architecture of the model (built using Python's **Keras** and TensorFlow libraries) will be based on Muriel Kosaka's speech emotion recognition model which utilizes the **Librosa** library for the Python programming language to extract the Mel spectrogram and obtain the cepstral data, subsequently tuning the **hyperparameters** (parameters with values that control a model's learning) to obtain the most optimized model.[26] The extraction process of this data can be seen in <u>Figure 2</u>, and the architecture of the model can be seen in <u>Figure 3</u> (see *Appendix B* for model diagram). The full list of functions is contained within *Appendix C*.[27]

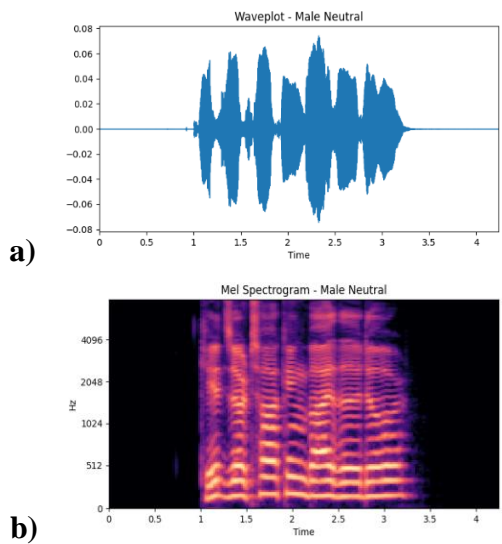

| Model: "sequential" | | |
|---|---|---|
| Layer (type) | Output Shape | Param # |
| conv1d (Conv1D) | (None, 250, 64) | 704 |
| conv1d_1 (Conv1D) | (None, 241, 128) | 82048 |
| max_pooling1d (MaxPooling1D) | (None, 40, 128) | 0 |
| dropout (Dropout) | (None, 40, 128) | 0 |
| conv1d_2 (Conv1D) | (None, 31, 128) | 163968 |
| max_pooling1d_1 (MaxPooling1 | (None, 5, 128) | 0 |
| dropout_1 (Dropout) | (None, 5, 128) | 0 |
| flatten (Flatten) | (None, 640) | 0 |
| dense (Dense) | (None, 256) | 164096 |
| dropout_2 (Dropout) | (None, 256) | 0 |
| dense_1 (Dense) | (None, 6) | 1542 |

Total params: 412,358
Trainable params: 412,358
Non-trainable params: 0

**Figure 2:** Audio extraction of a 'neutral' male voice as a waveform **(a)** and Mel spectrogram **(b)**

**Figure 3:** EmotioNN model architecture as CNN (Keras)

After training the model on all the voice actor audio clips to a user-defined number of **epochs** (evolutions where the dataset is split into training and testing portions to be evaluated before beginning the next epoch), the NN will be prepared to predict new audio files outside of the training set. The best model is saved only when an epoch has found a new maximum accuracy (improvement) score, to which it will retain the training knowledge and become more likely to accurately classify new data.

For the vocal isolator, the algorithm will be based upon the **REPET-SIM** method[28] with slight modifications, including utilizing a smaller FFT window overlap and converting non-local filters into soft masks using Wiener filtering.[29] The audio (WAV) file will be converted into a spectrogram, apply a filter to aggregate and constrain similar audio frames, reduce the bleed of the vocal and instrumental/accompaniment (or noise) masks, then separate the masks into background (noise) and foreground (voice) spectrums. Once these spectrums are separated, the foreground will contain the newly filtered audio with the vocals – which can then be exported to a new audio file.

## 5. Evaluation

The final model was trained to 2,000 epochs, achieving approximately 73% accuracy from the test data (see *Appendix B*). The **loss** from the model – also known as the cost or objective function, used to find the best parameters, known as **weights** for the model – was much greater in the test validation than the training validation; while accuracy scaled relatively proportionally between testing and training (see <u>Figure 4a</u>), loss diverged away from minimization compared to the training model and continued to grow exponentially over time (see <u>Figure 4b</u>).

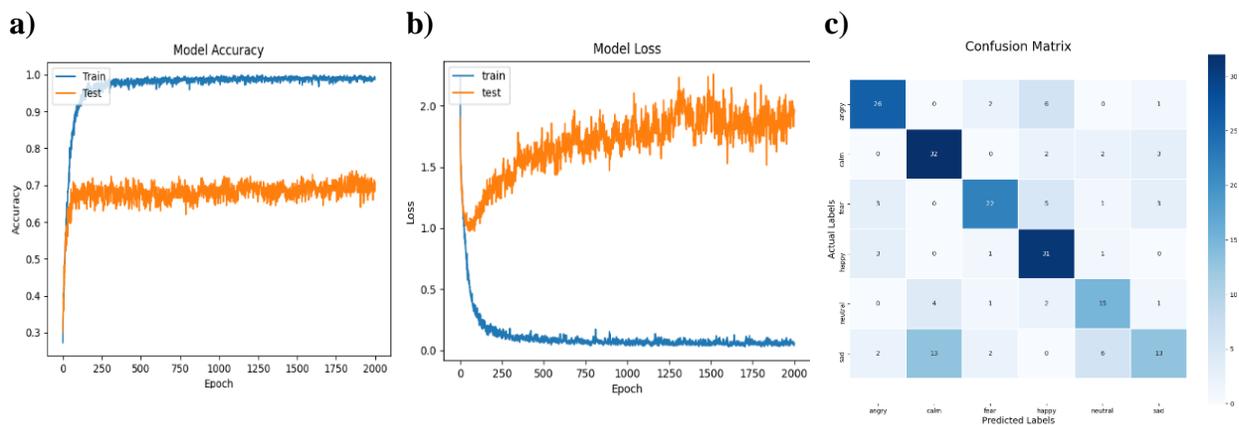

**Figure 4:** Final trained model after 2,000 epochs: accuracy **(a)**, loss **(b)**, and confusion matrix **(c)**

While this may appear problematic, an entropy of approximately 1.8 out of six classes gives a loss of 0.3, or roughly 30% – inverse to the accuracy value. To see the actual loss based on how the model predicts classes of emotions, a **confusion matrix** can be used to compare actual versus predicted labels (see Figure 4c). While this level of accuracy is decent, a much more complex architecture is necessary to train a better performing model in the future, especially including training on acoustic properties among the MFCCs.

## 6. Discussion

While the accuracy of the model may appear questionable, most results are surprisingly correct – albeit subjectively in some instances. *Appendix A* contains the data analyzed from having the CNN attempt to classify audio through multiple user-defined splits of a given file (in seconds), rather than testing the overall emotion of the sound. Tests were performed with and without isolating the vocals from the accompaniment of various art songs, including one choral piece and one purely instrumental piece; however, making these changes did not appear to make a visible difference in how the model predicted various segments of a song. Vocals, both isolated and accompanied, showed nearly identical results so long as the segment contains a voice – silence was typically the only unidentical factor (seen especially in *Appendix A, Sections 1 and 5*), and this may be for one of two reasons in particular: either the silence contained residual harmony (even minuscule) from the accompaniment or noise and the isolated version of the file did not, or the file is comprised of just the accompaniment which became silent after isolating the vocals (also seen in *Appendix A, Sections 3 and 4*). As well, changing the length of the segments also did not appear to make a difference in the perceived emotion either, as the smaller splits only added to the precision of the model's accuracy (see *Appendix A, Section 2.1*). Of course, as expected, the model did not fare well attempting to classify emotions of instrumental music given its lack of training and more precise **psychoacoustical** parameters (see *Appendix A, Section 4*).

One major flaw of the model appears to be its evolutionary progression; once the model achieves approximately 70% accuracy, it seldom improves. In an attempt to train the CNN 10,000 epochs, the model eventually failed after over 7500 epochs and its accuracy flatlined at less than 15% accuracy and never recovered, though the loss continued to change over time (see <u>Figure 5a</u> and <u>5b</u>). Strangely, these weights were saved as the "new best" by the model anyway, and the CNN appeared to predict "angry" for every given piece of data (see <u>Figure 5c</u>).

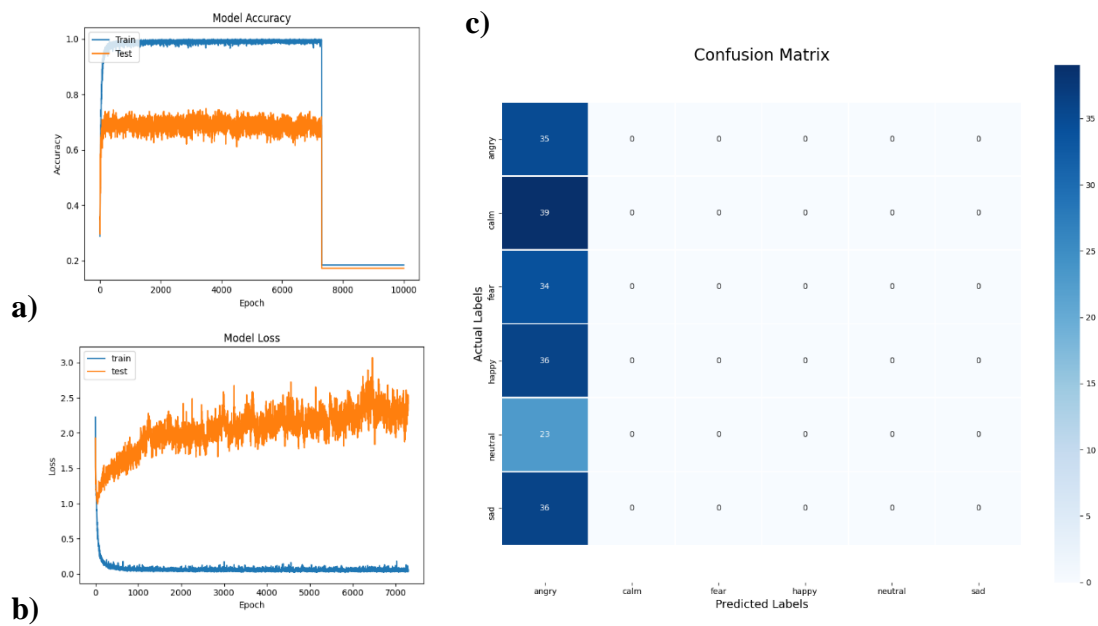

**Figure 5:** Early trained model to 10,000 epochs: accuracy **(a)**, loss **(b)**, and confusion matrix **(c)**

Additionally, in the final model, classifying in real-time also causes the fully trained (73% accurate) model to predict "angry" as well, but classifying these files later on yields normal results.

## 7. Conclusion

This paper focused on the creation of a CNN with the ability to accurately recognize emotions in the singing voice using MFCCs as the primary feature, as well as laying out the developmental plans for a future mobile (or desktop) application that utilizes this type of model for providing biofeedback. By training the model on pure vocal audio, the model was able to create a working classification memory even when data is noisy or includes instrumental accompaniment. Ideally, the model should be nearly or just as accurate without vocal isolation as with it – this appeared to be the case even with the current final model, fortunately.

In the future, this type of model and architecture could be expanded upon and utilized as the backend system of a much more complex piece of software – hopefully used in a biofeedback app as intended, providing visual feedback of both real-time and historical data for use in private music lessons. With finer tuning and the inclusion of more acoustic properties, this model has the potential to become much more accurate in a more refined architecture. Hopefully, a larger sung-emotion dataset will be created as well soon. Eventually, more training may be done on the analysis of choral music and, when a dataset permits, analyzing instrumental music both individually and for ensembles.

With a much more expansive and precise dataset, a future model could also be trained to recognize a wider array of emotions in both singing and speaking voices, as well as recognizing emotions in the breath, language (text – through **natural language processing**), face, and potentially body language, as a means of creating a near-true neural network amygdala. Given a more accurate and reliable, this technology could serve voice teachers and professionals alike as a means of training emotional expression with live feedback – especially in the case of rehearsing a very expressive work such as an aria or art song; with a reliable model, a teacher may have a mobile application which provides live feedback during a lesson or ensemble rehearsal, or an individual singer may use the biofeedback during practice sessions.

Appendix A: Spectral Analysis and Emotion Recognition Records of Segmented Data[30]

**1. Allerseelen – Richard Strauss (split every 20 seconds)**[31]

| Non-Isolated Vocals | | Isolated Vocals | |
|---|---|---|---|
| 0 | sad | 0 | sad |
| 20 | calm | 20 | calm |
| 40 | calm | 40 | calm |
| 60 | calm | 60 | calm |
| 80 | calm | 80 | calm |
| 100 | calm | 100 | calm |
| 120 | calm | 120 | calm |
| 140 | calm | 140 | calm |
| 160 | sad | 160 | sad |
| 180 | **_sad_** | 180 | **_calm_** |
| 200 | angry | 200 | angry |

**Table 1:** Comparison of isolated versus non-isolated vocals split into 20-second-long segments

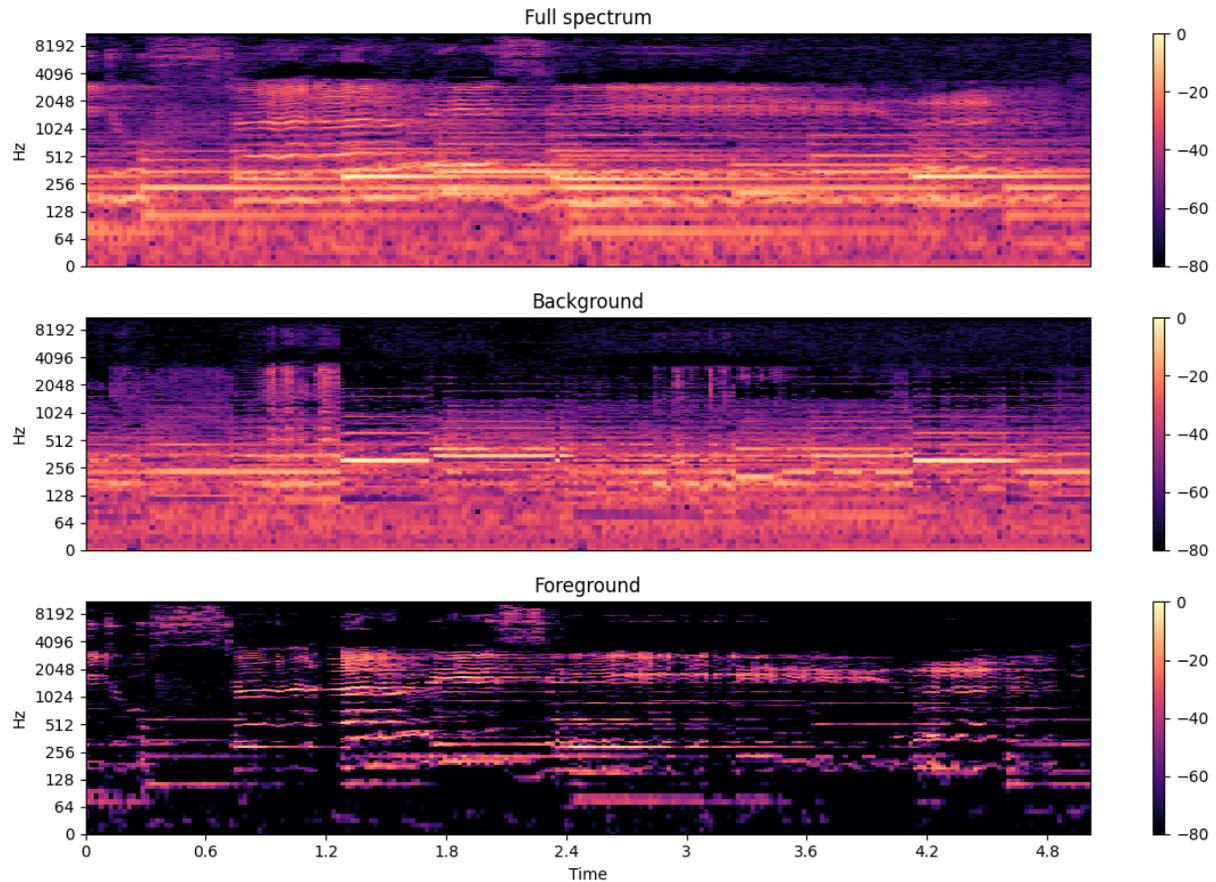

**Figure 6:** Spectral analysis of a baritone voice singing "Allerseelen" by Strauss (5 seconds)

## 2. Lenski's Aria – Pyotr Ilyich Tchaikovsky (split every 20 seconds)[32]

| Non-Isolated Vocals | | Isolated Vocals | |
|---:|---|---:|---|
| 0 | sad | 0 | sad |
| 20 | **calm** | 20 | **neutral** |
| 40 | calm | 40 | calm |
| 60 | calm | 60 | calm |
| 80 | calm | 80 | calm |
| 100 | calm | 100 | calm |
| 120 | calm | 120 | calm |
| 140 | **calm** | 140 | **sad** |
| 160 | calm | 160 | calm |
| 180 | calm | 180 | calm |
| 200 | calm | 200 | calm |
| 220 | **calm** | 220 | **happy** |
| 240 | calm | 240 | calm |
| 260 | **calm** | 260 | **sad** |
| 280 | **calm** | 280 | **fear** |
| 300 | **calm** | 300 | **fear** |
| 320 | calm | 320 | calm |
| 340 | calm | 340 | calm |
| 360 | sad | 360 | sad |
| 380 | angry | 380 | angry |

**Table 2:** Comparison of isolated versus non-isolated vocals split into 20-second-long segments

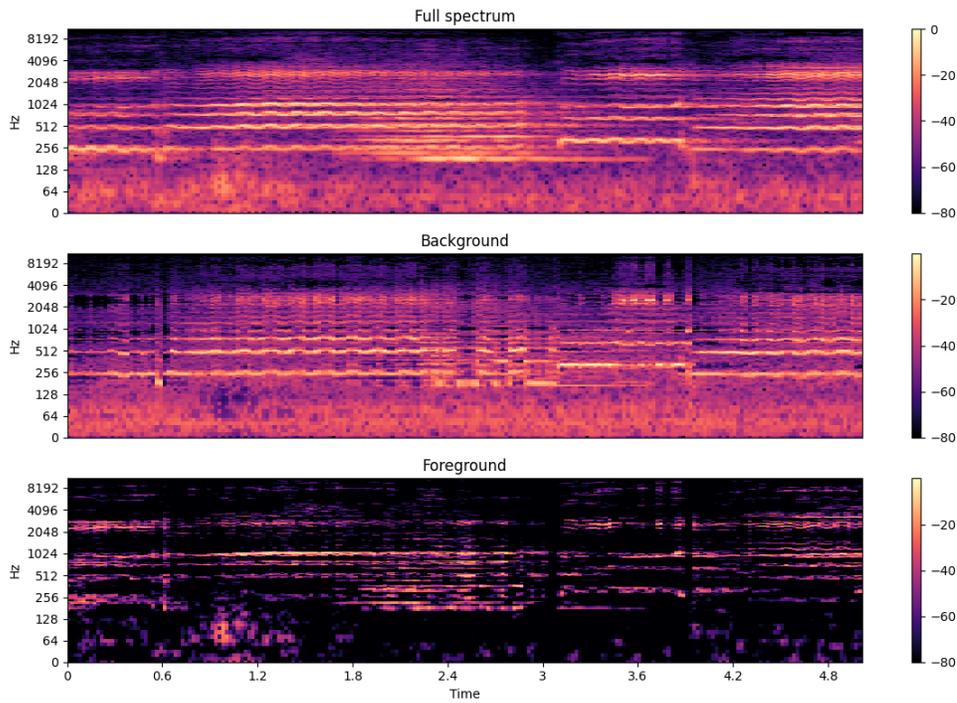

**Figure 7:** Spectral analysis of a tenor voice singing Lenski's Aria by Tchaikovsky (5 seconds)

**2.1. Lenski's Aria – Pyotr Ilyich Tchaikovsky (isolated vocals, split comparison)**

| Split 10 Seconds | | Split 20 Seconds | | Split 40 Seconds | | Split 60 Seconds | | Split 120 Seconds | |
|---|---|---|---|---|---|---|---|---|---|
| 0 | sad | 0 | sad | 0 | calm | 0 | sad | 0 | calm |
| 10 | sad | 20 | neutral | 40 | calm | 60 | calm | 120 | happy |
| 20 | sad | 40 | calm | 80 | calm | 120 | angry | 240 | sad |
| 30 | sad | 60 | calm | 120 | happy | 180 | sad | 360 | calm |
| 40 | calm | 80 | calm | 160 | happy | 240 | happy | | |
| 50 | happy | 100 | calm | 200 | calm | 300 | angry | | |
| 60 | calm | 120 | calm | 240 | calm | 360 | sad | | |
| 70 | calm | 140 | sad | 280 | fear | | | | |
| 80 | calm | 160 | calm | 320 | sad | | | | |
| 90 | sad | 180 | calm | 360 | calm | | | | |
| 100 | calm | 200 | calm | | | | | | |
| 110 | calm | 220 | happy | | | | | | |
| 120 | calm | 240 | calm | | | | | | |
| 130 | calm | 260 | sad | | | | | | |
| 140 | calm | 280 | fear | | | | | | |
| 150 | calm | 300 | fear | | | | | | |
| 160 | happy | 320 | calm | | | | | | |
| 170 | happy | 340 | calm | | | | | | |
| 180 | calm | 360 | sad | | | | | | |
| 190 | calm | 380 | angry | | | | | | |
| 200 | calm | | | | | | | | |
| 210 | calm | | | | | | | | |
| 220 | happy | | | | | | | | |
| 230 | calm | | | | | | | | |
| 240 | calm | | | | | | | | |
| 250 | happy | | | | | | | | |
| 260 | calm | | | | | | | | |
| 270 | fear | | | | | | | | |
| 280 | fear | | | | | | | | |
| 290 | happy | | | | | | | | |
| 300 | fear | | | | | | | | |
| 310 | calm | | | | | | | | |
| 320 | sad | | | | | | | | |
| 330 | happy | | | | | | | | |
| 340 | calm | | | | | | | | |
| 350 | calm | | | | | | | | |
| 360 | calm | | | | | | | | |
| 370 | calm | | | | | | | | |
| 380 | angry | | | | | | | | |

**Table 3:** Comparison of isolated vocals split into various segment lengths

### 3. Sure on this Shining Night – Morten Lauridsen (split every 20 seconds)[33]

| Non-Isolated Vocals | | Isolated Vocals | |
|---:|---|---:|---|
| 0 | calm | 0 | calm |
| 20 | **calm** | 20 | **happy** |
| 40 | calm | 40 | calm |
| 60 | calm | 60 | calm |
| 80 | sad | 80 | sad |
| 100 | calm | 100 | calm |
| 120 | happy | 120 | happy |
| 140 | **sad** | 140 | **fear** |
| 160 | angry | 160 | angry |
| 180 | calm | 180 | calm |
| 200 | calm | 200 | calm |
| 220 | **calm** | 220 | **sad** |
| 240 | calm | 240 | calm |
| 260 | **calm** | 260 | **sad** |

**Table 4:** Comparison of isolated versus non-isolated vocals split into 20-second-long segments

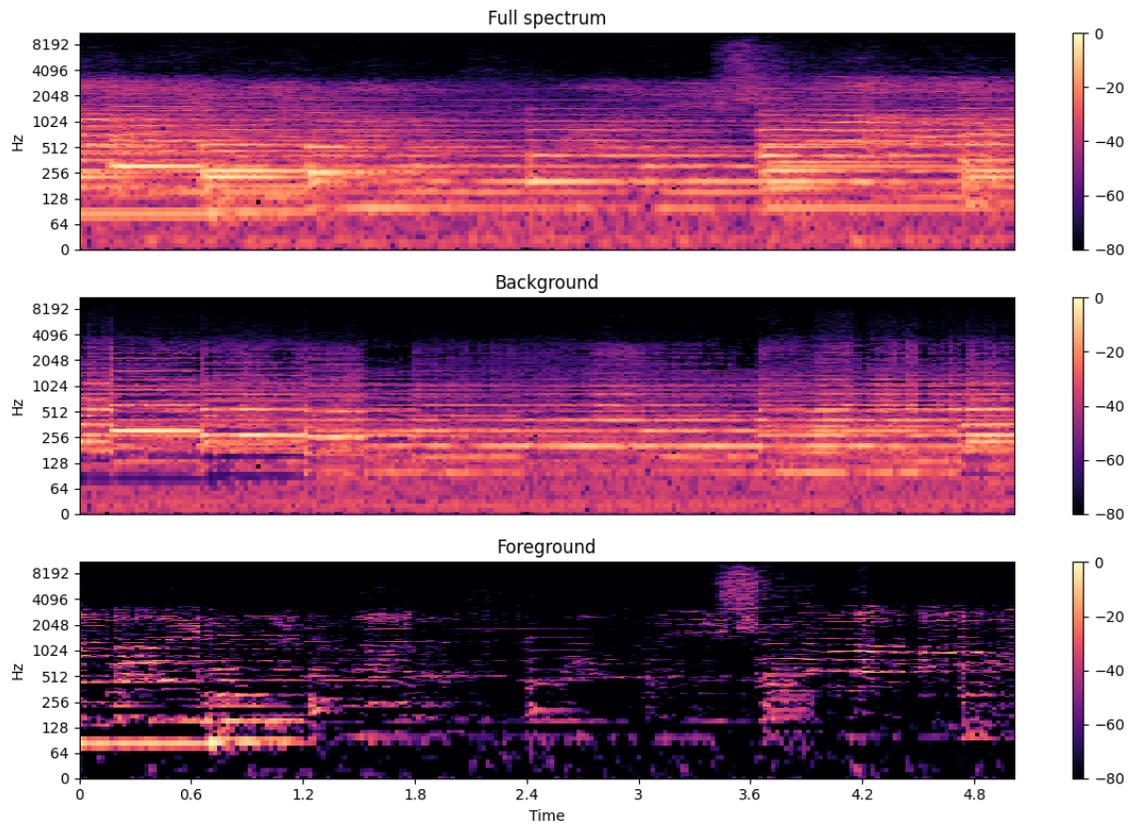

**Figure 8:** Spectral analysis of an SATB choir singing "Sure on this Shining Night" by Lauridsen (5 seconds)

**4. Coriolan Overture – Ludwig van Beethoven (split comparison)**[34]

| Orchestral Split 20 Seconds | | Orchestral Split 40 Seconds | |
|---|---|---|---|
| 0 | calm | 0 | calm |
| 20 | calm | 40 | calm |
| 40 | calm | 80 | calm |
| 60 | calm | 120 | calm |
| 80 | calm | 160 | calm |
| 100 | calm | 200 | calm |
| 120 | calm | 240 | calm |
| 140 | calm | 280 | calm |
| 160 | calm | 320 | calm |
| 180 | calm | 360 | calm |
| 200 | calm | 400 | calm |
| 220 | calm | 440 | sad |
| 240 | calm | 480 | calm |
| 260 | calm | | |
| 280 | calm | | |
| 300 | calm | | |
| 320 | calm | | |
| 340 | calm | | |
| 360 | calm | | |
| 380 | calm | | |
| 400 | calm | | |
| 420 | neutral | | |
| 440 | calm | | |
| 460 | calm | | |
| 480 | calm | | |

**Table 5:** Comparison of instrumental piece split into segments 20- versus 40-seconds long

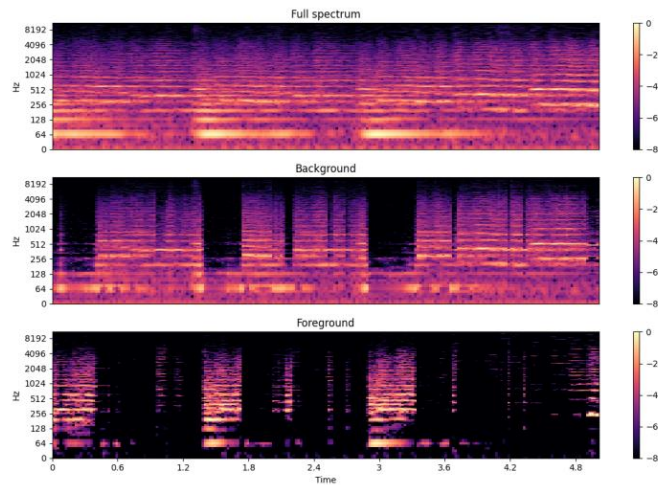

**Figure 9:** Spectral analysis of an orchestra performing the Coriolan Overture by Beethoven (5 seconds)

**5. Der Erlkönig – Schubert (split every 10 and 20 seconds)**[35]

| Non-Isolated Vocals (10 seconds) | | Isolated Vocals (10 seconds) | | Non-Isolated Vocals (20 seconds) | | Isolated Vocals (20 seconds) | |
|---|---|---|---|---|---|---|---|
| 0 | calm | <u>0</u> | calm | 0 | calm | 0 | calm |
| 10 | **<u>calm</u>** | <u>10</u> | **<u>sad</u>** | 20 | **<u>sad</u>** | 20 | **<u>fear</u>** |
| 20 | **<u>calm</u>** | <u>20</u> | **<u>sad</u>** | 40 | calm | 40 | calm |
| 30 | calm | <u>30</u> | calm | 60 | sad | 60 | sad |
| 40 | calm | <u>40</u> | calm | 80 | **<u>calm</u>** | 80 | **<u>sad</u>** |
| 50 | **<u>calm</u>** | <u>50</u> | **<u>sad</u>** | 100 | calm | 100 | calm |
| 60 | **<u>calm</u>** | <u>60</u> | **<u>sad</u>** | 120 | calm | 120 | calm |
| 70 | **<u>calm</u>** | <u>70</u> | **<u>sad</u>** | 140 | **<u>sad</u>** | 140 | **<u>calm</u>** |
| 80 | **<u>calm</u>** | <u>80</u> | **<u>sad</u>** | 160 | calm | 160 | calm |
| 90 | calm | <u>90</u> | calm | 180 | calm | 180 | calm |
| 100 | calm | <u>100</u> | calm | 200 | angry | 200 | angry |
| 110 | **<u>fear</u>** | <u>110</u> | **<u>happy</u>** | 220 | **<u>happy</u>** | 220 | **<u>angry</u>** |
| 120 | calm | <u>120</u> | calm | 240 | **<u>fear</u>** | 240 | **<u>calm</u>** |
| 130 | calm | <u>130</u> | calm | | | | |
| 140 | calm | <u>140</u> | calm | | | | |
| 150 | happy | <u>150</u> | happy | | | | |
| 160 | calm | <u>160</u> | calm | | | | |
| 170 | **<u>calm</u>** | <u>170</u> | **<u>sad</u>** | | | | |
| 180 | **<u>neutral</u>** | <u>180</u> | **<u>sad</u>** | | | | |
| 190 | happy | <u>190</u> | happy | | | | |
| 200 | angry | <u>200</u> | angry | | | | |
| 210 | **<u>happy</u>** | <u>210</u> | **<u>calm</u>** | | | | |
| 220 | happy | <u>220</u> | happy | | | | |
| 230 | **<u>calm</u>** | <u>230</u> | **<u>sad</u>** | | | | |
| 240 | **<u>fear</u>** | <u>240</u> | **<u>calm</u>** | | | | |

**Table 6:** Comparison of isolated versus non-isolated vocals split into 10- and 20-second-long segments

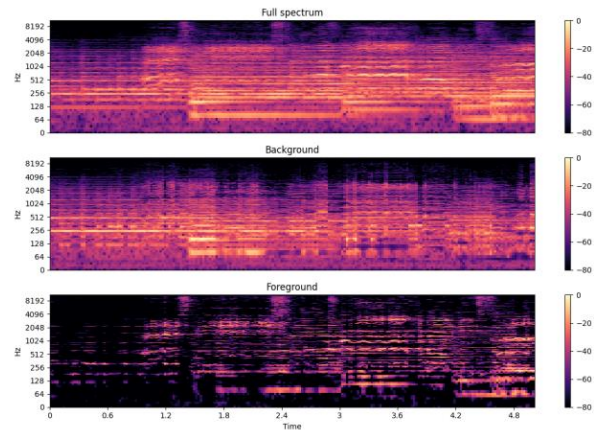

**Figure 10:** Spectral analysis of a baritone voice singing Der Erlkönig by Franz Schubert (5 sec.)

## Appendix B: Convolutional Neural Network Model and Performance

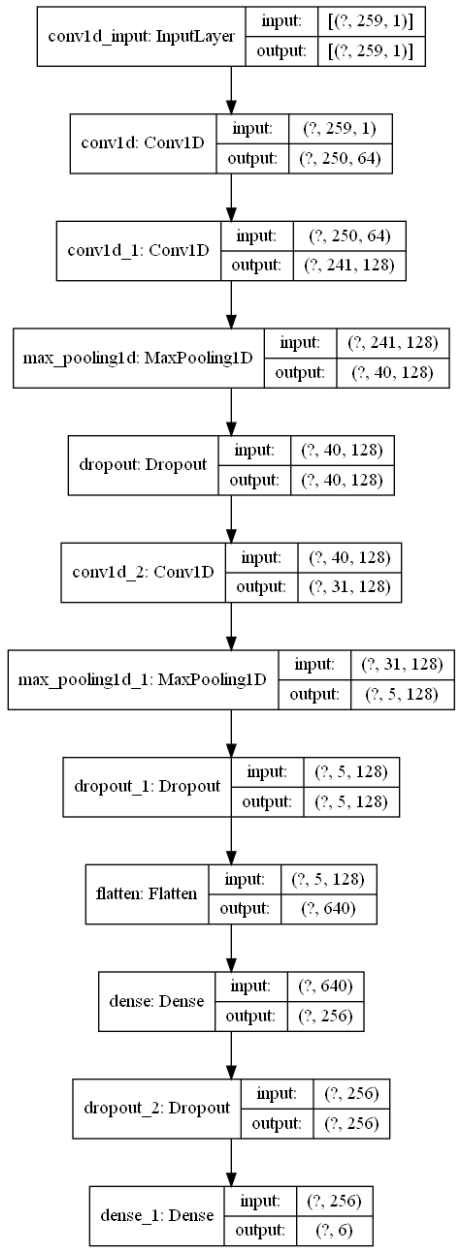

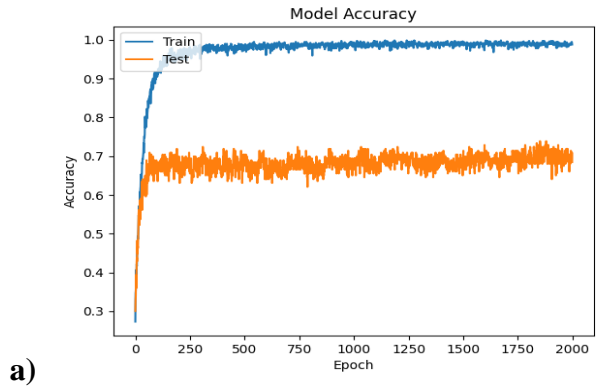

**a)**

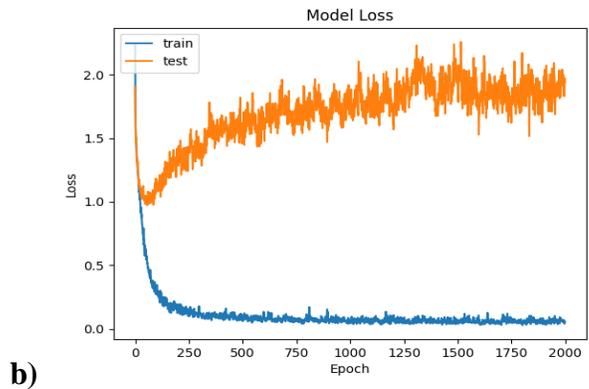

**b)**

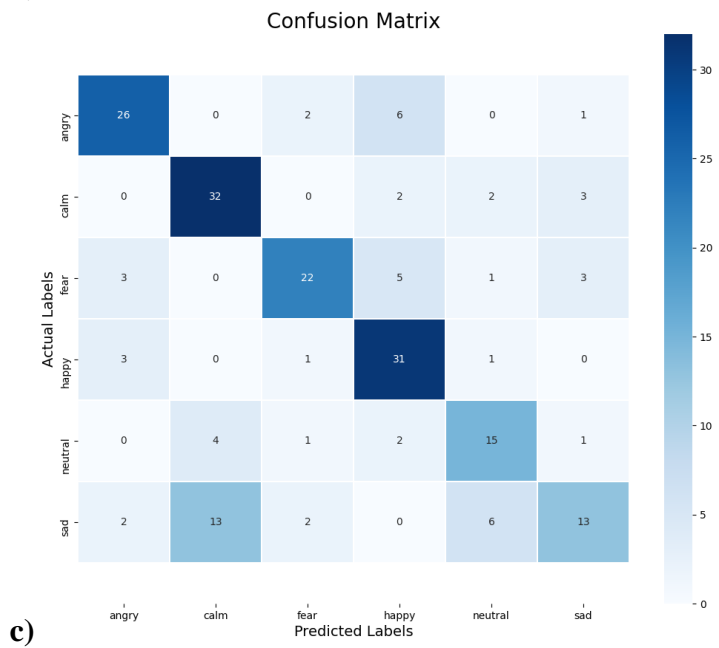

**c)**

**Figure 10:** Layer diagram of the EmotioNN model (input order)

**Figure 4:** Final trained model after 2000 epochs, accuracy **(a)**, loss **(b)**, and confusion matrix **(c)**

Appendix C: Backend System Architecture and Function Headers

The full implementation, including demonstrations and presentations can be found at: https://github.com/danielathome19/Sung-EmotioNN-Detector.

In *main.py*:

- **Libraries:**
  - Glob, math, time, threading, matplotlib, librosa, pyaudio, wave, IPython, numpy, pandas, sklearn, os, soundfile, sys, warnings, keras, tensorflow, seaborn, pydub
- **Classes:**
  - SplitWavAudio – Functions:
    - *get_duration():* Get duration of WAV file in seconds
    - *single_split(...):* Split a WAV file between two timestamps (in seconds)
    - *multiple_split(...):* Split a WAV file into multiple different splits of a user-defined duration (in seconds)
- **Functions:**
  - *demo(...):* Display a waveform and spectrogram of a WAV file
  - *make_classifier(...):* Build and return the CNN model given a dataset
  - *train():* Train the CNN model, plot accuracy and confusion, tune hyperparameters of model for higher accuracy
  - *predict(...):* Predict emotion of all files in a given folder
  - *record(...):* Record the user's microphone and save as a WAV file
  - *realtime(...):* Record the user's microphone and predict emotions every user-defined interval (in seconds)
  - *separatevocals(...):* Isolate vocals from WAV file, display mask spectrograms
  - *wavsplit(...):* Split WAV file into a user-defined duration (in seconds) using SplitWavAudio class
  - *main():* Function to act as a controller, calls any defined functions at runtime

Notes